\documentclass{gCFD2e}
\pdfoutput=1
\usepackage{multirow}
\usepackage{color}
\usepackage{amsmath}
\usepackage{amssymb}
\usepackage{amsthm}
\usepackage{caption}
\usepackage{subcaption}
\usepackage{epstopdf}
\usepackage[colorinlistoftodos,textsize=small]{todonotes}
\usepackage{moreverb}
\usepackage[colorlinks,bookmarksopen,bookmarksnumbered,citecolor=red,urlcolor=red]{hyperref}
\usepackage[percent]{overpic}
\usepackage{tikz}
\usetikzlibrary{arrows,shapes,patterns}
\usepackage{pgfplots}
\pgfplotsset{compat=1.8}

\usepackage{fnbreak}

\usepackage{color}
\definecolor{darkblue}{rgb}{0, 0, 0.5}
\definecolor{darkgreen}{rgb}{0, 0.5, 0}
\definecolor{darkred}{rgb}{0.5, 0.0, 0}
\definecolor{shaded}{gray}{.6}

\newcommand{\new}[1]{#1}

 \newcommand{\commentGeneric}[2]{\fbox{\begin{minipage}{0.6\textwidth}
 \textbf{#1:}~
 \begin{minipage}{0.8\textwidth}
 \it #2
 \end{minipage}\end{minipage}}\quad}
\newboolean{SHOWCOMMENTS}
\setboolean{SHOWCOMMENTS}{false}
\ifthenelse{\boolean{SHOWCOMMENTS}}{%
\newcommand{\commentU}[1]{\commentGeneric{Uli}{#1}}
}{
\newcommand{\commentU}[1]{}
}

\renewcommand{\vec}[1]{\mathbf{#1}}
\newcommand{\ia}[0]{\ensuremath{\alpha}}
\newcommand{\ib}[0]{\ensuremath{\beta}}

\renewcommand{\x}[0]{\ensuremath{\vec{x}}}
\renewcommand{\u}[0]{\ensuremath{\vec{u}}}
\renewcommand{\c}[0]{\ensuremath{\vec{c}}}
\renewcommand{\l}[0]{\ensuremath{\lambda}}

\DeclareMathOperator{\eq}{eq}
\DeclareMathOperator{\nq}{neq}
\DeclareMathOperator{\Rey}{Re}

\DeclareMathOperator{\Vol}{Vol}
\DeclareMathOperator{\Eo}{Eo}
\DeclareMathOperator{\Mo}{Mo}

\graphicspath{ {./figures/} }

\usepackage[capitalize]{cleveref}
\Crefname{equation}{Equation}{Equations}
\crefname{equation}{Eq.}{Eqs.}
\Crefname{section}{Section}{Sections}
\crefname{section}{Sec.}{Secs.}
\Crefname{definition}{Definition}{Definitions}
\crefname{definition}{Def.}{Defs.}
\Crefname{figure}{Figure}{Figures}
\crefname{figure}{Fig.}{Figs.}
\Crefname{table}{Table}{Tables}
\crefname{table}{Tab.}{Tabs.}
\Crefname{appsec}{Appendix}{Appendices}
\crefname{appsec}{appendix}{appendices}
\Crefname{proposition}{Proposition}{Propositions}
\crefname{proposition}{Prop.}{Props.}
\Crefname{theorem}{Theorem}{Theorems}
\crefname{theorem}{Theorem}{Theorems} 
\crefname{property}{}{}

\begin{document}


\title{Direct simulation of liquid-gas-solid flow with a free surface lattice Boltzmann method}

\author{Simon Bogner$^{\rm a}$\thanks{$^\ast$Corresponding author. Email: s.bogner@fz-juelich.de
\vspace{6pt}}, Jens Harting$^{\rm a, \rm b}$ and Ulrich R{\"u}de$^{\rm c}$\\\vspace{6pt} 
$^{a}${\em{Forschungszentrum J\"ulich,
  Helmholtz-Institut Erlangen-N\"urnberg f{\"u}r Erneuerbare Energien,
  F\"urther Stra{\ss}e  248,
  90429 N\"urnberg,
  Deutschland}};
$^{b}${\em{Faculteit Technische Natuurkunde, Technische Universiteit Eindhoven, P.O. Box 513, 5600 MB Eindhoven, Nederland}};
$^{c}${\em{Lehrstuhl f\"ur Systemsimulation, 
  Universit\"at Erlangen-N\"urnberg,
  Cauerstra{\ss}e 11,
  91058 Erlangen,
  Deutschland
}}
}

\maketitle

\begin{abstract}

Direct numerical simulation of liquid-gas-solid flows is uncommon due to the
considerable computational cost. As the grid spacing is determined by the
smallest involved length scale, large grid sizes become necessary -- in
particular if the bubble-particle aspect ratio is on the order of 10 or
larger. Hence, it arises the question of both feasibility and reasonability. In
this paper, we present a fully parallel, scalable method for direct numerical
simulation of bubble-particle interaction at a size ratio of 1-2 orders of
magnitude that makes simulations feasible on currently available super-computing
resources. With the presented approach, simulations of bubbles in suspension
columns consisting of more than $100\,000$ fully resolved particles become
possible. Furthermore, we demonstrate the significance of particle-resolved
simulations by comparison to previous unresolved solutions. The results indicate
that fully-resolved direct numerical simulation is indeed necessary to predict
the flow structure of bubble-particle interaction problems correctly.


\end{abstract}


\commentU{Hervorhebungen durch quotation marks vs. italic font - vereinheitlichen.}

\section{\label{sec:introduction}Introduction}
Due to the computational complexity of fluid-solid and liquid-gas-solid flow
problems, numerical solutions are usually based on homogenized models
\citep{Pan2016,Panneerselvam2009,Li2015}. Homogenized models do not resolve all
involved scales and model the phase interaction based on closure relations (drag
correlations) instead. The closure relations, in turn, are obtained from
experiments, or -- with the advent of high-speed computers -- by direct
numerical simulation (DNS) of systems of smaller size. \new{DNS techniques allow the most accurate predictions by resolving even the smallest relevant length scales. For fluid-solid, particulate flows the smallest typical length scale is the particle diameter.}
For the case of
fluid-solid flows, new drag correlations have been derived from numerical data
\citep{Beetstra2007,Tenneti2011,BognerMohanty,Tang2015}. Also, DNS has helped to
investigate the behavior of particle suspensions and to study hydrodynamic
interaction in particulate flows \citep{Aidun2010,Tenneti2014} in full detail.
\new{Due to computational costs, the system sizes that can be realized by DNS are limited compared to unresolved and homogenized models. Nevertheless, DNS is an important tool that allows the study of flow structures in every detail, and provides the most accurate solutions.}

To date, only few direct numerical simulation models for liquid-gas-solid
flows (LGS) can be found in literature. A numerical method for this class of
flows must combine a two-phase flow solver \citep{Scardovelli99,Tryggvason2011}
with a structural solver for the suspended solid
phase. 
Most LGS simulation approaches
\citep{Li1999b,Chen2004,vanSintAnnaland2005,Xu2013,Sun2015,Li2015} do not fully
resolve the particle geometry within the flow. This means that hydrodynamic
interaction between particles cannot be captured fully in these models\new{ which thus do not count as DNS models according to the narrower definition applied here}. Nevertheless,
these approaches make use of discrete particle methods
\citep{Bicanic2004,DeenEtAl2007} to resolve particle-particle collisions. The
first models to resolve both bubble and particle geometries have been presented
by \citet{Deen2009} and \citet{Baltussen2013}. These \new{DNS }models combine a
front-tracking liquid-gas method with an immersed boundary approach
\citep{Mittal2005} to couple the flow with the particle simulation. Recently,
the same group applied their methodology to study the effective drag on bubbles
and particles in liquid flow \citep{Baltussen2017}. Presumably due to
computational limitations, the simulated systems contain bubbles and particles
of similar size only. In many situations of practical relevance, however, the
particle size is much smaller than the bubble size. Alternatively, there are
efforts to combine diffusive multiphase models with particle models
\citep{Stratford2005,Jansen2011,JoshiSun}. These \new{DNS }models work with fully resolved
particles, but additional limitations arise from the necessity to resolve also
the liquid-gas interface -- especially at a high density difference. \new{The density ratio is on the order of $\mathcal{O}(10)$ in these models, which is much smaller than the density ratio of most liquid-gas two-phase flows. Only recently, \citet{Connington2015} have reached high density ratios for special cases.}

In the following, we present a \new{DNS }model for liquid-gas-solid flow that
allows the simulation of bubble-particle interaction in containing liquid. The
model is based on the free surface lattice Boltzmann method (FSLBM) of
\citep{KoernerEtAl} for high liquid-gas density ratios combined with the
particulate flow model of \citep{Ladd1994a,Ladd2001}. Based on a previous effort
\citep{Bogner2013}, we have developed a model that is inherently parallel and
allows bubble sizes one order of magnitude larger than the particle size while
still fully resolving the single particle geometries. \new{Since the grid
  spacing must be smaller than the particle diameter, the total number of
  lattice sites is necessarily large and the computational cost is
  considerable. Therefore, } the model is implemented based on a parallel
software framework \citep{walberla2011}, that has already been used to realize
massively parallel simulations of suspensions \citep{Goetz2010} and bubbly flows
\citep{Donath2009}. The new liquid-gas-solid model enables detailed studies of
particle transport in the wake of rising bubbles. We demonstrate that our model
is capable of predicting important suspension properties, such as increased
effective viscosity with solid volume fraction correctly. The terminal rise
velocity of a gas bubble decreases accordingly in simulations. Furthermore, we
validate the free surface model for different bubble regimes according to the
classification of Grace (spherical, ellipsoidal, skirted, dimpled), and present
examples of particle transport and mixing in the wake of a single rising bubble
for the different regimes. \new{The cost of the DNS is considerable. However, a
  comparison of the results to previous unresolved simulations indicates the
  necessity of DNS to predict the full characteristics of the flow and the
  induced particle transport.}


\section{\label{sec:method}Method}
In the following, we use a hybrid method based on the free surface lattice
Boltzmann method (FSLBM) of \citet{KoernerEtAl} and the particulate flow
model of \citet{Ladd1994a,Ladd2001}.
The computational domain is subdivided into the three disjoint regions,
corresponding each to the space occupied by liquid, gas, or solid phase,
respectively.

\subsection{Hydrodynamic Lattice Boltzmann Model}
To solve the hydrodynamic equations for the liquid region, we use a D$3$Q$19$
lattice Boltzmann model \citep{Wolf-Gladrow,QianEtAl1992} on a Cartesian
grid. The lattice velocities are denoted by $\c_q$ with $q=0, \dots, 18$ and
have units of grid spacing $\delta_x$ per time step $\delta_t$. The data $f_q$
with $q=0, \dots, 18$ of the scheme is also called \emph{particle distribution
  function} (PDF).  The \emph{lattice Boltzmann equation} of the model can be
written as,
\begin{subequations}
  \begin{align}
    f_q(\x +\c_q \delta_t, t+\delta_t) &= f^{*}_q(\x,t),  \label{eq:LBE1} \\
    f_q^{*}(\x,t) &= f_q(\x, t) + \l_- f_q^{\nq, -} + \l_+ f_q^{\nq,+},  \label{eq:LBE2}
  \end{align}
\end{subequations}
where $f^*(\x,t)$ has been substituted, and is referred to as the
\emph{post-collision} state. \new{The upper-indices ``$+$/$-$'' denote the
  even/odd parts of the respective function.} The right hand side of
\cref{eq:LBE2} corresponds to the two relaxation time collision operator of
\citet{Ginzburg2007}, with the odd and even eigenvalues $\l_-, \l_+ \in
(-2,0)$. These eigenvalues thus control the relaxation of the even and odd parts
of the non-equilibrium,
\begin{equation}
  f^{\nq}_q(\x, t) = f_q(\x,t) - f_q^{\eq}(\x,t),
\end{equation}
defined as the deviation from the equilibrium function
$f_q^{\eq} = e_q( \rho(\x,t), \u(\x,t) )$, given as the polynomial
\begin{equation}
  e_q(\rho, \u) = \rho w_q \left( 1 + \frac{c_{q,\ia} u_\ia}{ c_s^2 } + \frac{u_\ia u_\ib}{ 2 c_s^4 } ( c_{q, \ia}c_{q, \ib} - c_s^2 \delta_{\ia \ib} ) \right),
  \label{eq:equilibrium}
\end{equation}
where the $w_q$, $q=0, \dots, 18$, are a set of lattice weights, and the
constant $c_s = \delta_x / (\sqrt{3} \, \delta_t)$ is called the \emph{lattice
  speed of sound}. The \emph{macroscopic} flow variables of pressure and
velocity are moments of the PDF, that is,
\begin{subequations}
  \begin{align}
    p(\x,t) = c_s^2 \rho(\x,t) &= c_s^2 \sum_{q=0}^{18}{ f_q (\x,t) },\\
    u_{\ia}(\x,t) &= \frac{1}{\rho} \sum_{q=0}^{18}{ c_{q,\ia} f_q (\x,t) }.
  \end{align}
\end{subequations}
It can be shown that the velocity field is a second order accurate solution to
the incompressible Navier-Stokes equations \citep{Frisch87,Holdych2004,Junk2005}
with kinematic viscosity
\begin{equation}
  \nu = - \left( \frac{1}{\lambda_{+}} + \frac{1}{2} \right) c_s^2  \delta_t.
  \label{eq:latticeViscosity}
\end{equation}
While the first relaxation parameter $\lambda_+$ is chosen according to the
desired flow viscosity, the second parameter $\lambda_-$ is fixed to satisfy the
equation,
\begin{equation}
  \left( \frac{1}{\lambda_{+}} + \frac{1}{2} \right) \left( \frac{1}{\lambda_{-}} + \frac{1}{2} \right) = \frac{3}{16}.
\end{equation}
This ``magic'' parameterization  is optimal for straight axis
aligned wall boundaries \citep{Ginzburg1994}, and yields viscosity independent
solutions in general geometries \citep{Ginzburg2009}.

\subsection{\label{sec:FSLBM}Free Surface Lattice Boltzmann Method (FSLBM)}
\begin{figure}
  \centering
  \begin{tikzpicture}[scale=0.666]
    \tikzstyle{interface} = [fill=lightgray,font=\footnotesize,inner sep=1pt]

    \draw[fill=gray] (0,0) rectangle (1,6);
    \draw[fill=gray] (1,3) rectangle (2,6);
    \draw[fill=gray] (2,4) rectangle (3,6);
    \draw[fill=gray] (3,5) rectangle (9,6);
    \draw[fill=gray] (8,0) rectangle (9,5);
    \draw[fill=gray] (6,0) rectangle (8,1);
    \draw[fill=gray] (7,1) rectangle (8,2);

    \draw[fill=lightgray] (1,1) rectangle (2,3);
    \draw[fill=lightgray] (2,2) rectangle (3,4);
    \draw[fill=lightgray] (3,3) rectangle (4,5);
    \draw[fill=lightgray] (4,4) rectangle (7,5);
    \draw[fill=lightgray] (1,0) rectangle (6,1);
    \draw[fill=lightgray] (5,1) rectangle (7,2);
    \draw[fill=lightgray] (6,2) rectangle (8,3);
    \draw[fill=lightgray] (7,3) rectangle (8,5);

    \draw[help lines, black] (0,0) grid (9,6);
    \begin{scope}[scale=1.5]
      \begin{scope}[rotate = 30]
        \draw (3.4,0.0) ellipse (2 and 1.2);
      \end{scope}
      \node[interface] at (4.3,3.0) {$I(t)$};
    \end{scope}
  \end{tikzpicture}
  \hspace{1.0cm}
  \begin{tikzpicture}[scale=0.666]
    \draw[fill=gray] (0,0) rectangle (1,1); \node[anchor=south west] at (1.5,0) {liquid cells};
    \draw[fill=lightgray] (0,2) rectangle (1,3); \node[anchor=south west] at (1.5,2) {interface cells};
    \draw[fill=white] (0,4) rectangle (1,5); \node[anchor=south west] at (1.5,4) {gas cells (inactive)};
  \end{tikzpicture}
  \caption{\label{fig:cellTypes}A fictitious interface $I(t)$ and its discrete
    FSLBM representation consisting of gas, interface, and liquid nodes.}
\end{figure}
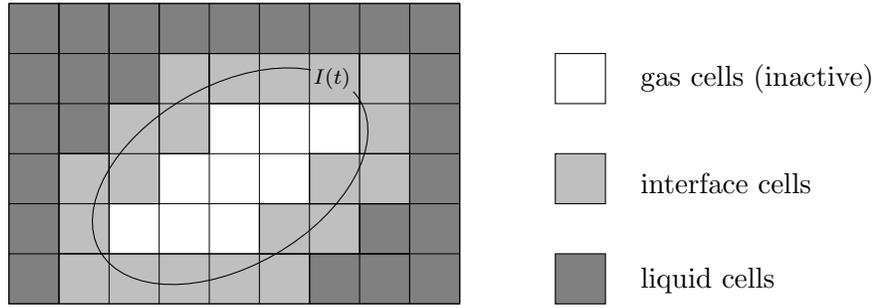

The FSLBM is an interface capturing scheme that is based on the \emph{volume of
  fluid} \citep{Hirt81,Tryggvason2011} approach. The \emph{fill level} or
\emph{volume fraction} $\varphi(\x)$ serves as indicator function. For each node
$\x$, the fill level $\varphi(\x)$ is defined as the volume fraction of liquid
within the cubic cell volume around $\x$. \Cref{fig:cellTypes} shows that three
different types of cells can be distinguished:
\begin{itemize}
\item $C_g(t)$: the set of \emph{gas nodes}, where $\varphi = 0$.
\item $C_l(t)$: the set of \emph{liquid nodes}, where $\varphi = 1$.
\item $C_i(t)$: the set of \emph{interface nodes}, where $0 < \varphi \leq
  1$. An interface node $\x$ always has a liquid neighbor $\x + \delta_t \c_q
  \in C_l$ and a gas neighbor $\x + \delta_t \c_p \in C_g$ for some
  $p,q=1,\dots,18$.
\end{itemize}
The nodes in $C_l \cup C_i$ are active lattice Boltzmann nodes. Introducing
further the set of obstacle nodes $C_s(t)$ that are not part of the fluid domain
(e.g., walls or nodes that are blocked out by particles), $C_g \cup C_s$ forms
the set of inactive nodes within the simulation domain.

The set of interface nodes $C_i$ is exactly the set of boundary nodes that
possess a neighbor in the gas subdomain. If an interface node $\vec{x}_b \in
C_i$ has a gas neighbor $\vec{x} + \delta_t \c_q \in C_g$, then the boundary
condition of \citet{KoernerEtAl},
\begin{equation}
  f_{\bar{q}}(\x_b, t+1) = - f^{*}_q(\x_b,t) + 2 e_q^{+}(\rho_w, \vec{u}_w),
  \label{eq:fsk}
\end{equation}
is applied for the opposite direction $\bar{q}$ with $-\c_q=\c_{\bar{q}}$. Here,
$p_w = c_s^2 \rho_w$ defines the boundary value for pressure, and $\vec{u}_w$
represents the flow velocity at the boundary. It can be shown that \cref{eq:fsk}
yields a first order approximation of a free boundary \citep{BognerAmmer}.

The interface capturing scheme is updated according to the flow simulation in
every time step.  The indicator function $\varphi(\x,t)$ is updated directly
from the lattice Boltzmann data. \begin{subequations}
  \begin{equation}
    \varphi(\x_i, t+1) = \varphi(\x_i,t) + \frac{1}{\rho(\x_i,t+1)} \left( \sum_{q=1}^{Q-1} \Delta m_q(\x_i,t)  \right),
    \label{eq:massExchange}
  \end{equation}
  with the direction-dependent exchange mass
  \begin{equation}
    \small
    \Delta m_q(\x_i,t) = 
    \begin{cases} 
      0 & \text{ if } \x_i+\c_q \notin (C_i \cup C_l),\\
      \frac{1}{2}(\varphi(\x_i+\c_q) + \varphi(\x_i)) (f_{\bar{q}}(\x_i +\c_q) - f_q(\x_i)) & \text{ if } \x_i+\c_q \in C_i,\\
      f_{\bar{q}}(\x_i +\c_q) - f_q(\x_i) & \text{ if } \x_i+\c_q \in C_l,
    \end{cases}
    \label{eq:massExchangeb}
  \end{equation}
\end{subequations}

\begin{figure}
  \centering
  \begin{tikzpicture}[scale=1.0]
    \matrix[nodes={draw, thick},
    row sep=0.3cm,column sep=2.0cm] {
      \node[rectangle,fill=gray,minimum width=2.5cm,minimum height=1.2cm] (liquid) {liquid} ;&
      \node[rectangle,fill=lightgray,minimum width=2.5cm,minimum height=1.2cm] (interface) {interface} ;&
      \node[rectangle,minimum width=2.5cm,minimum height=1.2cm] (gas) {gas};\\
    };

    \draw[->,-latex,thick,bend angle=35,bend right,name=itol] (interface) to node[midway,above]{\small{$\varphi(t+\delta_t) \geq 1.0 
        $}} (liquid) ;
    \draw[->,-latex,thick,bend angle=35,bend left,name=itog] (interface) to node[midway,above]{\small{$\varphi(t+\delta_t) \leq 0.0 
        $}} (gas);
    \draw[->,-latex,thick,bend angle=35,bend right,name=ltoi] (liquid) to node[midway,below,text width=3.1cm,align=center]{\small{neighbor converts into gas}} (interface);
    \draw[->,-latex,thick,bend angle=35,bend left,name=gtoi] (gas) to node[midway,below,text width=3.1cm,align=center]{\small{neighbor converts into liquid}} (interface);
  \end{tikzpicture}
  \caption{\label{fig:cellConversions}Possible cell state conversions in FSLBM
    simulations. Conversions of interface cells are triggered by the fill level
    $\varphi$ and, for gas and liquid cells, by conversions of neighboring
    interface cells into liquid or gas, respectively. \citep{myDiss}}
\end{figure}
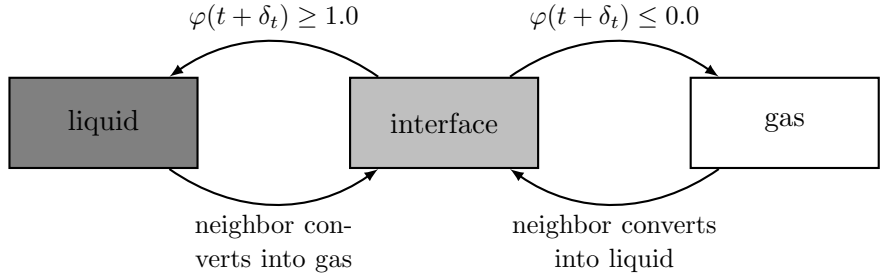

The sets $C_l$, $C_i$, and $C_g$ are updated according to the rules illustrated
in \cref{fig:cellConversions}, where each arrow corresponds to a possible state
transition: The transitions between gas, liquid, and interface state, are
triggered according to the fill levels $\varphi$ of the interface
cells. Whenever the fill level of an interface cell becomes equal to 0 (equal to
1), then a conversion into a gas cell (liquid cell) is triggered.  If needed,
inverse transitions from liquid (gas) into interface state are performed in
order to close the layer of interface cells.
If a gas node changes to interface state, then its LBM data is initialized based
on the equilibrium, \cref{eq:equilibrium}. Details can be found in
\citet{KoernerEtAl} and in \citep{Bogner2013} for moving particles.

Since the flow of the gas phase is not simulated in the free surface model, a
special treatment of the individual bubbles, i.e., connected regions of gas, is
necessary \citep[cf.][]{Anderl2014a,Caboussat2005,KoernerEtAl}. Such a
\emph{bubble model} conserves the mass in the gas phase and provides the local
gas pressure $p_g$ needed to define the boundary condition \cref{eq:fsk}.


For the simulation of capillary flows, the Laplace pressure jump across the
interface can be included in \cref{eq:fsk}. The boundary value for the pressure
is then
\begin{equation}
  p_w = p_g(\x,t) + 2 \sigma \kappa(\x,t),
\end{equation}
where $p_g$ is the pressure in the gas bubble and $\kappa$ is the local
curvature of the interface. Following \citet{Brackbill92}, the curvature can be
computed from the fill levels, based on the equations,
\begin{subequations}
  \begin{equation}
    \vec{n} = \nabla \varphi ,
  \end{equation}
  \begin{equation}
    \kappa(\x) = - ( \nabla \cdot \hat{\vec{n}} ).
  \end{equation}
\end{subequations}
To evaluate these expressions, we use an optimized finite difference scheme
according to \citet{ParkerYoungs}. At the solid particles, perfect wettability
is assumed. Further details and alternative curvature computation schemes can be
found in \citet{Cummins2005,Popinet2009,BognerHarting}.

\subsection{Particulate Flow Simulation}
Each (spherical) particle is defined by its radius $R_P$, specific density
$\rho_P$, and a Lagrangian description consisting of position $\x_P(t)$,
velocity $\u_P(t)$, and angular velocity $\vec{ \omega }_P(t)$. Any grid node
inside of a particle is called obstacle node. Whenever a liquid node $\x_b$ is
next to an obstacle node covered by particle $P$, then the \emph{bounce-back}
rule with velocity term,
\begin{equation}
  f_{\bar{q}}(\x_b, t+1) = f_q^*(\x_b,t) - 2 e_q^-(\rho_w, \u_w),
  \label{eq:bbrule}
\end{equation}
is used to impose the particle surface velocity,
\begin{equation}
  \u_w = \u_P + \vec{ \omega }_P \times ( \x_w - \x_P ),
\end{equation}
at the boundary. In \cref{eq:bbrule}, $\rho_w$ is substituted with the density
value of the boundary point from the previous time step. 

For the time integration of the particle data, the hydrodynamic force
$\vec{F}_P$ and torque $\vec{T}_P$ are computed from the lattice Boltzmann
data. Based on the \emph{momentum exchange principle}, one computes
\begin{align}
  \vec{F}_P &= \sum_{\vec{x} \in B_P}{ \sum_{q \in I_P(\vec{x})} { \Delta \vec{j}_q(\vec{x})  \frac{\delta_x^3}{\delta_t} } } \label{eq:netforce} ,\\
  \vec{T}_P &= \sum_{\vec{x} \in B_P}{ \sum_{q \in I_P(\vec{x})}{ (\vec{x} -
      \vec{x}_P) \times \Delta \vec{j}_q(\vec{x}) \frac{\delta_x^3}{\delta_t} } } \label{eq:torque},
\end{align}
where
\begin{equation}
  \Delta \vec{j}_q(\x) := \c_q f^*(\x,t) - \c_{\bar{q}} f(\x, t+1),
  \label{eq:momentum_transfer}
\end{equation}
is used to approximate the momentum transferred along a single
boundary-intersecting link at a boundary node $\x$
\citep{Ladd1994a,Ladd2001}. In \cref{eq:netforce,eq:torque}, the set $B_P$ is
the set of all nodes surrounding the particle $P$ with a nonempty set
$I_P(\x_b)$ of particle surface intersecting links. If $B_P$ contains lattice
nodes on the inside of another particle, the equilibrium distribution is assumed
in \cref{eq:momentum_transfer}.

The hydrodynamic lubrication forces obtained by \cref{eq:netforce} are valid
only if the gap between two particles $P_1$, $P_2$ is sufficiently
resolved. Hence, if the gap size becomes smaller than $\Delta_c = 2/3 \delta_x$,
then a \emph{lubrication correction},
\begin{equation}
  \vec{F}^{lub}_{{P_1},{P_2}} = - \frac{ 6 \pi \mu (R_{P_1} R_{P_2})^2 }{ (R_{P_1} + R_{P_2})^2 }   \left( \frac{ 1 }{ |\vec{x}_{1,2}| - R_{P_1} - R_{P_2} } - \frac{1}{\Delta_c} \right) \hat{\vec{x}}_{1,2} \cdot (\vec{u}_{P_1} - \vec{u}_{P_2}) \hat{\vec{x}}_{1,2},
  \label{eq:lubrication}
\end{equation}
is added to the net force $\vec{F}_{P_1}$, where $\x_{1,2} = \x_{P_2} -
\x_{P_1}$ is the relative position of the particles \citep{Ladd2001}. Here,
$\mu=\rho \nu$ is the dynamic viscosity. This improves the simulation of
hydrodynamic interaction between particles \citep{Aidun2010}. Since
\cref{eq:lubrication} diverges for $|\vec{x}_{1,2}| \rightarrow 0$, the gap size
is limited from below to be at least $0.2 \Delta_c$. Furthermore, to increase
stability at higher solid volume fractions, the time integration of the
particles proceeds in up to $10$ time steps per LBM step. 

\new{The inclusion of wetting boundaries is described in
  \citet{Brackbill92,BognerHarting}. However, we only study fully wetting
  particles in the following.}


\section{\label{sec:validation}Validation of the Numerical Model}
The simulations of bubbles in moderately dense suspensions and bubble-particle
interaction are found in \cref{sec:results}. Here, we first validate the correct
behavior of solid-liquid suspension simulations (\cref{sec:val:suspension}) and
gas-liquid simulations (\cref{sec:val:bubbles}) with our model.

\subsection{\label{sec:val:suspension}Validation of Particle Suspension Model}
\begin{figure}
  \begin{subfigure}[t]{0.49\textwidth}
    \centering
    \resizebox {\columnwidth} {!} {
      \begin{tikzpicture}[trim axis left]
        \begin{axis}[
          xlabel=solid volume fraction $\Phi$ \vphantom{Reynolds $\Rey_P$},
          ylabel=relative viscosity $\mu_s / \mu$,
          legend cell align=left,
          legend pos=north west,
          extra x ticks = {0.1},
          extra x tick labels = {\vphantom{$10^{-2}$}},
          ]
          \addplot[mark=x] table[x=Phi, y=heavy1] {./ViscosityPhi.csv};
          \addlegendentry{$\Rey_P=0.011$}
          \addplot[mark=o] table[x=Phi, y=heavy2] {./ViscosityPhi.csv};
          \addlegendentry{$\Rey_P=0.1$}
          \addplot[mark=square] table[x=Phi, y=heavy3] {./ViscosityPhi.csv};
          \addlegendentry{$\Rey_P=1$}
          \addplot[mark=triangle] table[x=Phi, y=heavy4] {./ViscosityPhi.csv};
          \addlegendentry{$\Rey_P=2$}

          \addplot[dashed,blue,line width=1.2,domain=0.08:0.42] { (1 + ((1.25*x)/(1 - x/0.63)) )^2 };
          \addlegendentry{\Cref{eq:Eilers}}
        \end{axis}
      \end{tikzpicture}
    }
    \caption{\label{fig:viscPhi}Relative viscosity at various solid volume
      fractions.}
  \end{subfigure}
  \begin{subfigure}[t]{0.49\textwidth}
    \centering
    \resizebox {\columnwidth} {!} {
      \begin{tikzpicture}[trim axis right]
        \begin{semilogxaxis}[
          xlabel=Reynolds number $\Rey_P$,
          xlabel near ticks,
          ylabel=relative viscosity $\mu_s / \mu$,
          legend cell align=right,
          legend pos=north west,
          ytick pos=right,
          ]
          \addplot[mark=x] table[x=Re, y=Phi10] {./ViscosityRe.csv};
          \addlegendentry{$\Phi=0.1$}
          \addplot[mark=o] table[x=Re, y=Phi20] {./ViscosityRe.csv};
          \addlegendentry{$\Phi=0.2$}
          \addplot[mark=square] table[x=Re, y=Phi30] {./ViscosityRe.csv};
          \addlegendentry{$\Phi=0.3$}
          \addplot[mark=triangle] table[x=Re, y=Phi40] {./ViscosityRe.csv};
          \addlegendentry{$\Phi=0.4$}     
        \end{semilogxaxis}
      \end{tikzpicture}
    }
    \caption{\label{fig:viscRe}Relative viscosity at various Reynolds numbers.}
  \end{subfigure}

  \caption{\label{fig:relativeVisc}Simulation of a thickening particle
    suspension in shear flow. The relative viscosity of the simulated suspension
    increases with solid volume fraction and Reynolds number. \citep{myDiss}}
\end{figure}
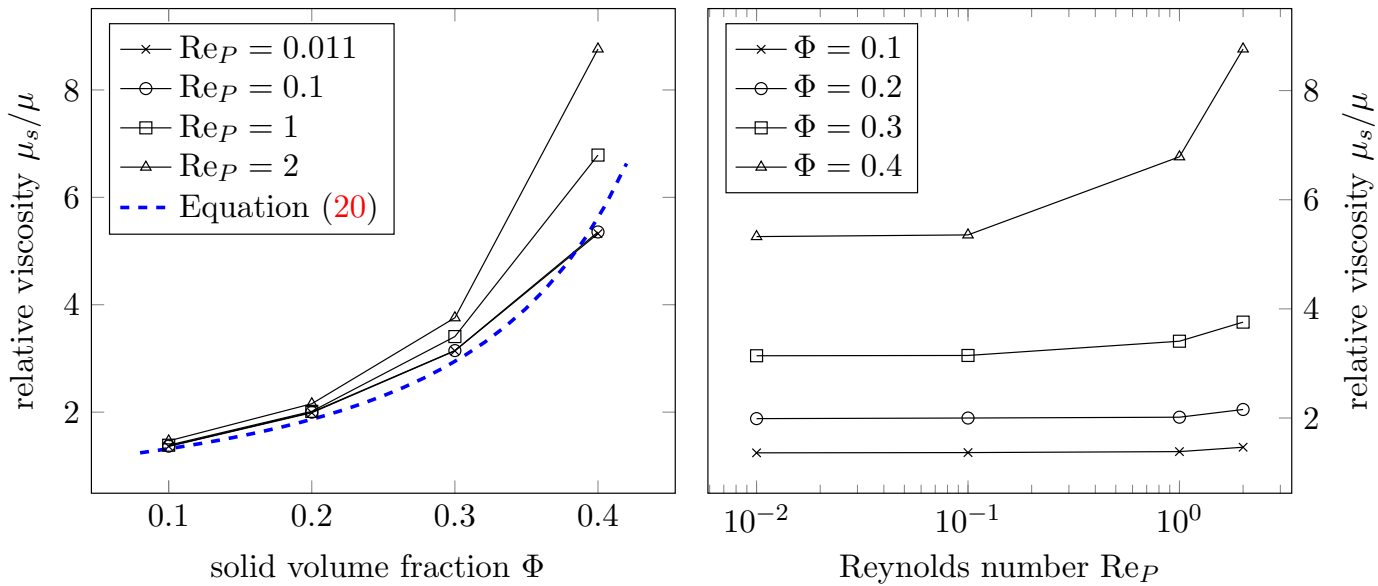

It has been demonstrated in the past that the LBM is valid in the simulation of
particle suspensions, e.g.,
\cite{Aidun2010,Ladd2001,Harting2014,Kromkamp2006}. Here, we reproduce as
validation experiment the relative shear viscosity of a spherical particle
suspension in a shear flow between plates.  Similar to \citet{Kromkamp2006}, a
domain is initialized with a random particle bed of $N$ spherical particles of
radius $R_P=8\delta_x$ and specific density $\rho_s=8$. The domain size is fixed
to $V=114\delta_x \times 116\delta_x \times 180\delta_x$, altering $N$ to
realize different solid volume fractions
\begin{equation}
  \Phi = \frac{N V_P}{V},
\end{equation}
where $V_P$ is the particle volume. The flow is initially at rest, and driven by
imposing a constant velocity $u_x = \pm 0.01 \delta_x/ \delta_t$ on the boundary
planes at $z=0$ and $z=180$ \new{in opposed directions}, while applying periodicity
along $x$ and $y$ directions. The effective viscosity $\mu_s$ of the numerical
suspension model is evaluated by measuring the net force $\bar{F}_x$ on the
boundary walls,
\begin{equation}
  \mu_s = \frac{ \bar{F}_x }{ A \dot{\gamma} },
\end{equation}
\new{where $\dot{\gamma}$ is the shear rate.}
The resulting force values oscillate due to non-trivial interaction between
particles, and must be averaged over a number of time steps $T$ (typically
$\dot{ \gamma } T \geq 100$). The (particle) Reynolds number is defined as
\begin{equation}
  \Rey_P = \frac{ \rho_f \dot{\gamma} (2 R_P)^2 }{\mu},
\end{equation}
where $\rho_f$ is the fluid density. In \citet{myDiss}, the same setup was
repeated with various solid volume fractions and Reynolds numbers. As shown in
\cref{fig:relativeVisc}, the model correctly predicts the expected increase of
effective viscosity with increased solid volume fraction and Reynolds
number. \Cref{fig:viscPhi} also displays the empirical correlation of Eilers
\citep{Stickel2005},
\begin{equation}
  \frac{\mu_s(\Phi)}{\mu} = \left[1 + \frac{1.25 \Phi}{1 - \Phi/\Phi_{\max}} \right]^2,
  \label{eq:Eilers}
\end{equation}
where $\Phi_{\max}=0.63$ is assumed as the maximal packing fraction for random
sphere packings.

\subsection{\label{sec:val:bubbles}Validation of Free Surface Model}
The FSLBM described in \cref{sec:FSLBM} has been validated for the case of
single rising bubbles in liquid columns in \citet{myDiss}, from which the
following results are adopted. 
The behavior of single rising bubbles in a quiescent (infinite) liquid is
characterized by three dimensionless numbers \citep{Clift78,FanTsuchiya}. The
\emph{Morton number} is defined as
\begin{equation}
  \Mo = \frac{g \mu^4 \Delta\rho}{\rho^2 \sigma^3},
  \label{eq:Morton}
\end{equation}
where $g$ is the gravitational constant, $\mu$ and $\rho$ are the viscosity
and density of the surrounding liquid, and $\Delta\rho$ is the density
difference of gas and liquid. 
The \emph{E{\"o}tv{\"o}s number},
\begin{equation}
  \Eo = \frac{\Delta\rho g d^2}{\sigma},
  \label{eq:Eotvos}
\end{equation}
where $d$ is the diameter of the bubble, characterizes the ratio of buoyancy and
surface tension forces. Finally, the bubble Reynolds number is defined as
\begin{equation}
  \Rey_b = \frac{\rho u_{\infty} d}{\mu},
\end{equation}
where $u_{\infty}$ is the terminal rise velocity of the bubble. The bubble
diameter $d$ is understood as the diameter of a volume-equivalent spherical
bubble, unless otherwise noted. Based on \citet{Grace73}, the behavior of gas
bubbles rising in a liquid column can be classified and allows a prediction of
the bubble shape, e.g., spherical, ellipsoidal, spherical cap, skirted, dimpled,
etc., or can be used to estimate the terminal rise velocity $u_\infty$ of the
bubble, if $\Eo$ and $\Mo$ are given. Alternatively, one can work with the
correlation of \citet{FanTsuchiya},
\begin{equation}
  \tilde{u}_{\infty} = \left[ \left( \frac{\Mo^{-1/4} \Eo}{K_b} \right)^{-n} + \left( \frac{2c}{\sqrt{\Eo}} + \frac{\sqrt{\Eo}}{2} \right)^{-n/2} \right]^{1/n},
  \label{eq:Fan}
\end{equation}
where $\tilde{u}_{\infty}$ is the nondimensional rise velocity. Velocity and
diameter are made nondimensional using
\begin{equation}
  \tilde{u} = u \left( \frac{\rho}{\sigma g} \right)^{1/4}, \text{ and } \tilde{d} = d \left( \frac{\rho g}{\sigma} \right)^{1/2}.
\end{equation}
In \cref{eq:Fan}, the parameters $K_b$, $c$, and $n$, are chosen to account for
special material properties not covered by $\Rey$, $\Mo$, and $\Eo$. The
parameter $n$ ranges from $0.8$ to $1.6$ depending on the liquid purity, while
$c$ is chosen as $1.2$ (single-component liquid) or $1.4$ (multi-component
liquid). The value of $K_b$ is adapted as
\begin{equation}
  K_b = \max(12, K_{b0} \Mo^{-0.038}),
  \label{eq:Kb}
\end{equation}
where $K_{b0}$ depends on the liquid (e.g., $K_{b0}=14.7$ for water). Like
the Grace diagram, correlation \cref{eq:Fan} is obtained from experimental data,
and can predict $u_\infty$ with an error of about $\pm 10 \%$.

\begin{table}
  \centering
  \resizebox {\columnwidth} {!} {
    \begin{tabular}{l | c c c | *{3}{c} | r }
      case               & $\mu$  $[Pa\, s]$& $\sigma$ $[\frac{N}{m}]$& $g$ $[\frac{m}{s^2}]$ & $\Mo$ & $\Eo$ & $\Rey_b$ & $\Rey_b^*$  \\
      \hline
      A (spherical)    &  0.25 &   0.145 & 0.981 & $1.26 \times10^{-3}$ & $0.974$ & $1.66$   & $1.76$     \\
      B (ellipsoidal)   &  0.42 &   0.145 & 9.81 & $0.100$                    & $9.74$  & $4.18$   & $4.43$     \\
      C (skirted)       &  0.13211 &   0.014545 & 9.81 & $0.9711$                   & $97.1$  & $18.56$ & $14.53$  \\
      D (dimpled)     &  0.75   &   0.014545 & 9.81 & $1018$                      & $97.4$  & $1.58$   & $1.46$    \\
    \end{tabular}
  }
  \caption{\label{tab:Annaland}Test cases for different bubble regimes. $\Rey_b$ is the expected Reynolds number for infinite domains according to \cref{eq:Fan} with $n=1.0$, $c=1.2$ and $K_{b0}=14$ in \cref{eq:Kb}. $\Rey_b^*$ is the value obtained in FSLBM simulations in a finite domain with free-slip walls. Simulations were parameterized according to the given viscosity $\mu$, liquid-gas surface tension $\sigma$, and gravitational constant $g$, assuming a liquid mass density of $\rho = 1000 kg / m^3$.}
\end{table}

\begin{figure}
  \centering
  \begin{tikzpicture}[scale=0.75]
    \fill[thick,lightgray] (0,0) rectangle (4,8);
    \fill[thick,white]  (2,2) circle (0.8);
    \fill[pattern=north east lines, line width=0] (0,-0.12) rectangle (4,0); 

    \draw[thick] (0,0) -- (4,0); \node[] at (2,-0.5) {no-slip};
    \draw[thick,dashed] (0,8) -- (4,8); \node[] at (2,8.5) {open (pressure)};
    \draw[thick] (0,0) -- (0,8); \node[rotate=-90] at (-0.5,4) {free-slip};
    \draw[thick] (4,0) -- (4,8); \node[rotate=+90] at (4.5,4) {free-slip};

    \draw[thick,<->] (1.2,2) -- (2.8,2); \node[] at (2.0,1.75) {$2R$};

     \node[] at (2.0,2.45) {$\Omega_g$};
     \node[] at (2.0,3.5) {$\Omega_l$};
  \end{tikzpicture}
  \caption{\label{fig:bubbleSetup}Boundary conditions around liquid column for
    rising bubble simulations. At $t=0$, a spherical bubble $\Omega_g$ of
    diameter $2R$ is initialized, surrounded by quiescent liquid $\Omega_l$.}
\end{figure}
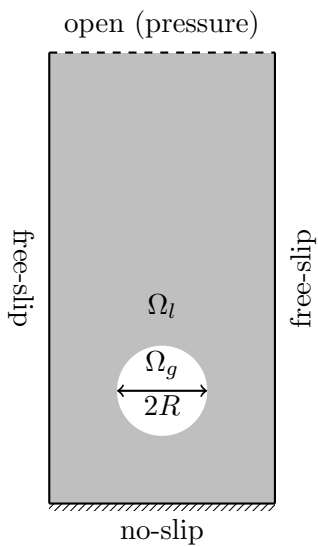

\Citet{Annaland2005} suggest four different bubble regimes for the validation of
a two-phase volume of fluid solver, that are used as a reference in the
following. The $\Mo$ and $\Eo$ numbers used in the following simulations
(\Cref{tab:Annaland}) have been chosen to represent the test cases suggested by
\citet{Annaland2005}. In this reference, the surface tension modeling is based
on \citet{Brackbill92}, similar to our finite difference model. The authors
suggest free-slip boundary conditions for all lateral directions, no-slip at the
bottom, and a pressure boundary at the top of the domain. A sketch of the domain
and initial conditions is shown in \cref{fig:bubbleSetup}.  The fluid parameters
for the simulation collected in \cref{tab:Annaland} are given with respect to a
liquid of density $\rho=1000 kg/m^3$. The grid spacing is $\delta_x = 10^{-3}m$
and the time step is $\delta_t = 10^{-4}s$. As a compromise between
computational cost and influence of the finite domain size on the bubble
dynamics, a domain size of $40 \times 40 \times 100$ nodes is recommended by the
authors. Here, we directly adopt the resolution and the domain size of the
original. The initial condition consists of a \emph{spherical} bubble of
$R=6\delta_x$ centered around the position $(20,20,10) \delta_x$ in a column of
quiescent liquid. 

\begin{figure}
  \centering
   \begin{subfigure}[t]{0.33\linewidth}
    \centering
    \includegraphics[width=\linewidth]{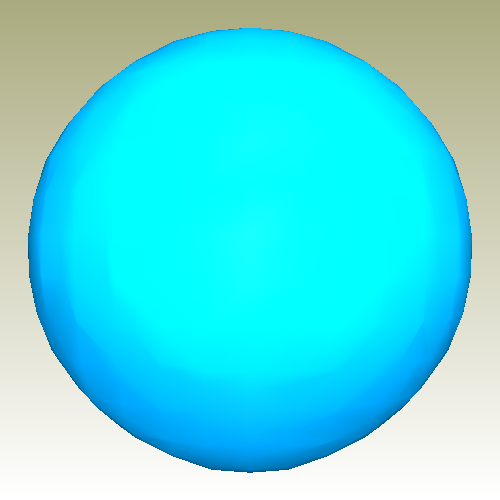}
    \caption{Case A (spherical), $t=20000\delta_t$}
  \end{subfigure}
  \begin{subfigure}[t]{0.33\linewidth}
    \centering
    \includegraphics[width=\linewidth]{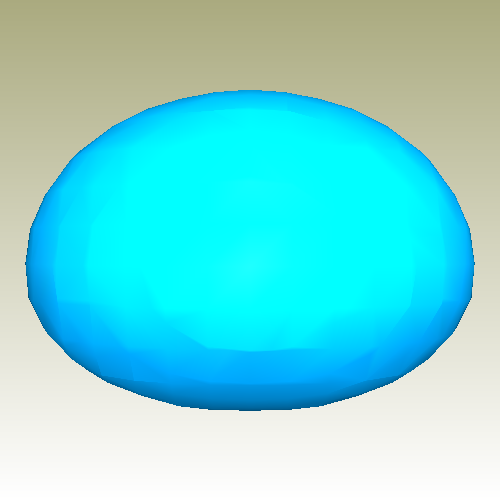}
    \caption{Case B (ellipsoidal), $t=5000\delta_t$}
  \end{subfigure}

  \begin{subfigure}[t]{0.33\linewidth}
    \centering
    \includegraphics[width=\linewidth]{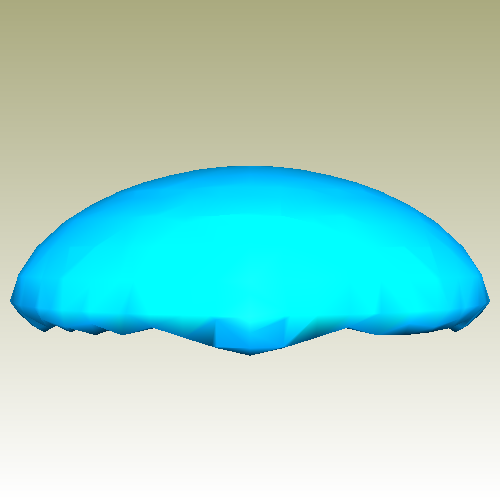}
    \caption{Case C (skirted), $t=4000\delta_t$}
  \end{subfigure}
  \begin{subfigure}[t]{0.33\linewidth}
    \centering
    \includegraphics[width=\linewidth]{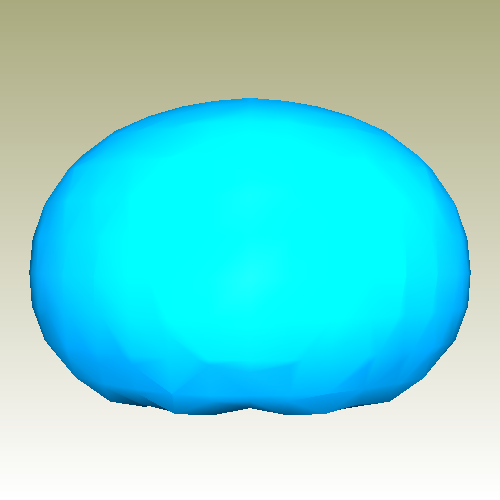}
    \caption{Case D (dimpled), $t=8000\delta_t$}
  \end{subfigure}
  
  \caption{\label{fig:shapes}Bubble shapes obtained from simulations of the
    bubble regimes from \cref{tab:Annaland}. }
\end{figure}
\begin{figure}
  \centering
   \begin{subfigure}[t]{0.2\linewidth}
    \centering
    \includegraphics[width=\linewidth]{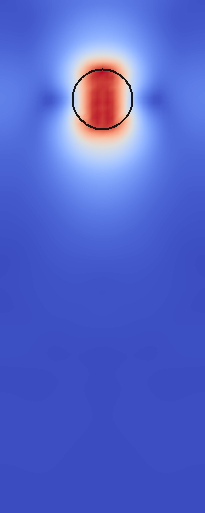}
    \caption{Case A (spherical), $t=20000\delta_t$}
  \end{subfigure}
  \begin{subfigure}[t]{0.2\linewidth}
    \centering
    \includegraphics[width=\linewidth]{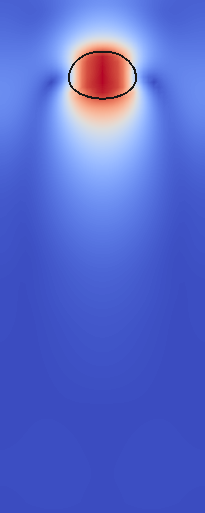}
    \caption{Case B (ellipsoidal), $t=5000\delta_t$}
  \end{subfigure}
  \begin{subfigure}[t]{0.2\linewidth}
    \centering
    \includegraphics[width=\linewidth]{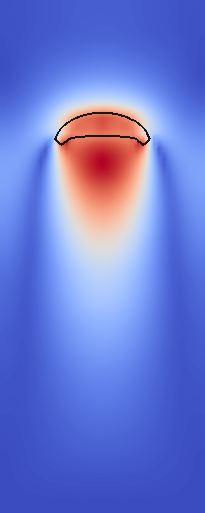}
    \caption{Case C (skirted), $t=4000\delta_t$}
  \end{subfigure}
  \begin{subfigure}[t]{0.2\linewidth}
    \centering
    \includegraphics[width=\linewidth]{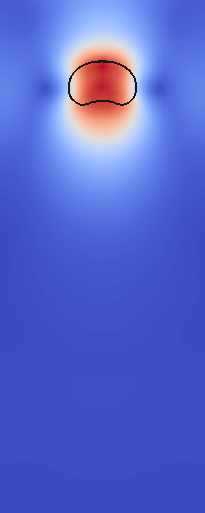}
    \caption{Case D (dimpled), $t=8000\delta_t$}
  \end{subfigure}

  \caption{\label{fig:velFields}Velocity field in the slice $y=20\delta_x$ of
    the simulations of the bubbles regimes from \cref{tab:Annaland}. Red color
    indicates high flow velocity, blue indicates low velocity magnitude. Black
    lines indicate the free surface.}
\end{figure}

For $t \geq 0$, the initially spherical bubble starts to accelerate due to the
pressure gradient until it reaches a terminal velocity and the bubble shape does
not change any more. \Cref{fig:shapes} shows the simulated bubble shapes using a
triangulation of the smoothed indicator function contour surface
$\varphi=0.5$. For each case, the simulated bubble shape agrees well with the
predictions according to the Grace diagram. The velocity field around the bubble
is shown in \cref{fig:velFields}. As reported also by \citet{Annaland2005}, the
terminal velocities obtained from simulations agree reasonably well with the
predictions of experimental relations. \Cref{tab:Annaland} lists the simulated
terminal Reynolds number $\Rey_b^*$ in comparison to the prediction according to
\cref{eq:Fan}.



\section{\label{sec:results}Results}
\subsection{\label{sec:bubble-particle}Bubble Particle Interaction}
We now study cases of bubble-particle interaction. Again, the basic setup
consists of a liquid column containing a single bubble of radius $R=0.01m$
rising within a bed of spherical particles of radius $R_P=8 \cdot 10^{-4}m$.
Using a grid spacing of $\delta_x=8 \cdot 10^{-5}m$, the size of the
computational domain is $500 \times 500 \times 1300$ lattice cells surrounding
the initially spherical bubble at $(250,250,250)\delta_x$. This means that each
particle is resolved by $5$ lattice cells per diameter.  The surface tension
$\sigma$ is $0.145N/m$ and the gravity is assumed to be $0.981 m/s^2$, such that
the spherical bubble regime is expected, in a fluid of density is $\rho =
1000 kg/m^3$ and viscosity $\mu=0.25 kg/(m s)$. Within the liquid column, a homogeneous particle bed is
initialized by choosing random positions. The bed density $\Phi$ varies with the
particle number $N$,
\begin{equation}
  \Phi = \frac{N V_P}{V - V_b},
\end{equation}
where $V_P = 4/3 \pi R_P^3$ is the particle volume and $V_b=4/3 \pi R^3$ is the
bubble volume. The solid mass density is $\rho_s= 3000 kg/m^3$. At the given
bubble-particle size ratio, the liquid-solid system surrounding the bubble can
be viewed as a homogeneous medium of increased density and
viscosity. \new{Notice, that the effective time scale of particle sedimentation
  is low compared to the expected rise velocity of the bubble.}

\begin{figure}
   %
    \centering
    \resizebox {0.5\columnwidth} {!} {
      \begin{tikzpicture}[trim axis left]
        \begin{axis}[
          xlabel=solid volume fraction $\Phi$ ,
          ylabel=terminal velocity $\bar{u}_{\infty}$,
          legend cell align=left,
          legend pos=north west,
          xticklabels = { 0, $0 \%$, $2 \%$, $4 \%$, $6 \%$, $8 \%$, $10 \%$},
          xticklabel style = {font=\small}
          ]
          \addplot[mark=x] table[row sep=\\] {%
            Phi          velocity\\
            0             0.53687069058757\\
            0.01         0.54376978332139\\
            0.03         0.52166135018381\\
            0.05         0.50675825176692\\
            0.075       0.48794013951139\\
            0.1           0.47890956185243\\
          };
        \end{axis}
      \end{tikzpicture}
    }
    \caption{\label{fig:bubbleParticle}Dimensionless terminal rise velocity at
      solid volume fractions from $0$ to $10\%$.}
\end{figure}
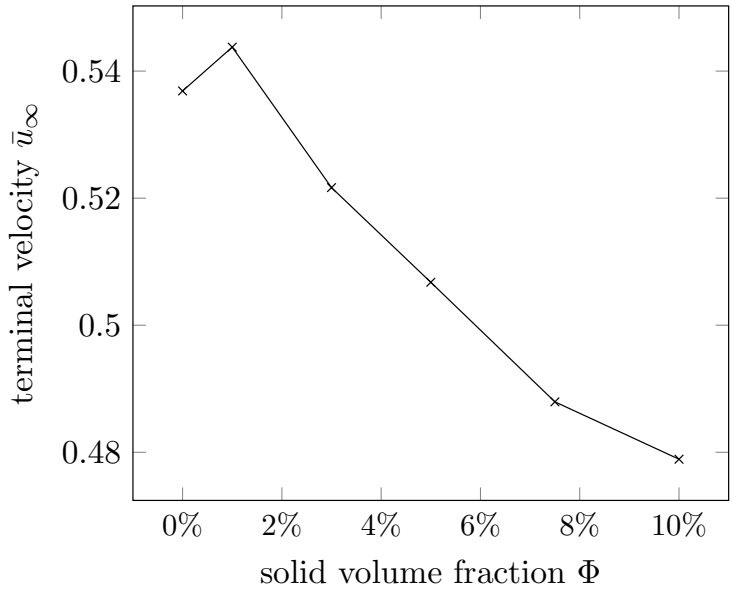

In a series of simulations the bubble is released within the particle bed, and
the terminal rise velocity depending on the bed solid volume fraction is
evaluated.  \Cref{fig:bubbleParticle} shows the decrease of bubble velocity with
increased bed density. Due to the presence of the particles, the average mass
density of the particle suspension around the bubble increases, and the buoyancy
force on the bubble is increased. This explains the small increase in velocity
from $0$ to $1 \%$ solid volume fraction. More significantly, with higher solid
volume fraction, the higher effective viscosity of the suspension reduces the
terminal velocity reached. This is in agreement with experiments from
literature, that reports a decreasing rise velocity with increased suspension
thickness \citep{Tsuchiya1997}.

We remark that the model is currently limited to low solid volume fractions. For
the spatial resolution applied in this work, solid volume fractions of $20 \%$
often developed instabilities and nonphysical behavior. The reason seems to be
the distribution of particles next to the liquid-gas interface that can make the
free-surface algorithm ineffective by covering the cells containing the
liquid-gas interface. \new{This is currently a limitation of the model, which
  might be improved by altering the surface tension model to satisfy the perfect
  wettability of particles more accurately.}

\subsection{\label{sec:mixing}Simulation of Bubble-Induced Particle Mixing}
The particle bed is now limited to the range $z=[0,500] \delta_x$ that includes
the initially spherical bubble.  We assume a liquid density of
$\rho=1000 kg/m^3$, viscosity $\mu=10^{-1}kg/(m \cdot s)$, liquid-gas surface tension
$\sigma = 0.1 N/m$, and a gravity $g=-9.81m/s^2$ along the $z$-axis. Choosing
the time step as $\delta_t = 2.5 \cdot 10^{-5}s$, the lattice relaxation time
becomes $\tau \approx 1.672$. The dimensionless numbers for the bubble are
$\Mo = 9.81 \cdot 10^{-4}$ and $\Eo = 39.24$, with an expected terminal Reynolds
number $\Rey_b=61.73$ according to \cref{eq:Fan} (with $n=1$, $c=1.2$). This
setup has been chosen similar to a test case of \citet{Deen2007BP}. 

\begin{figure}
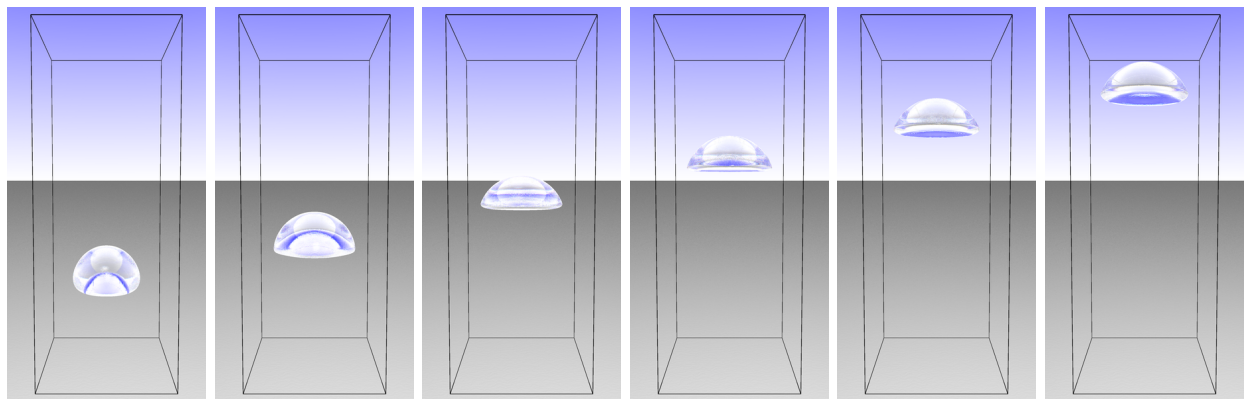
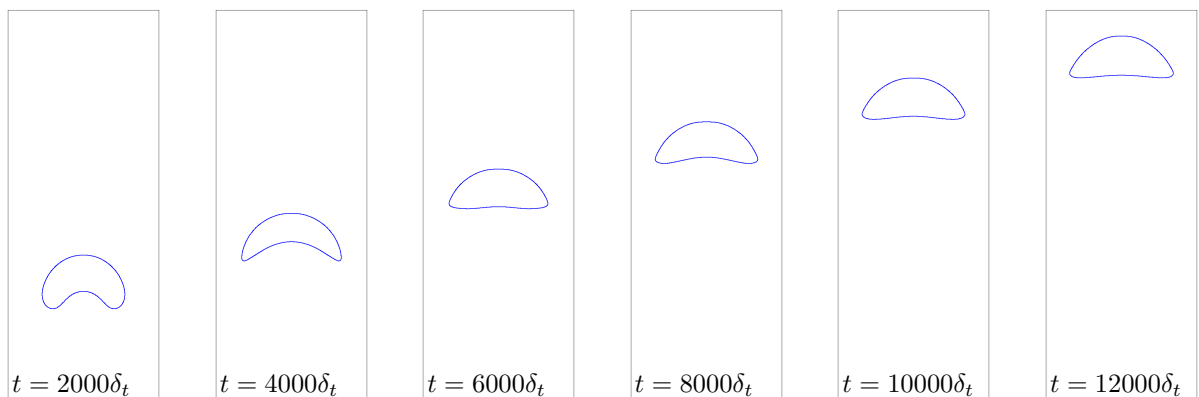

  \centering

  \begin{subfigure}[t]{\linewidth}
    \includegraphics[width=0.16\linewidth]{{{./Phi0_2000}}}
    \includegraphics[width=0.16\linewidth]{{{./Phi0_4000}}}
    \includegraphics[width=0.16\linewidth]{{{./Phi0_6000}}}
    \includegraphics[width=0.16\linewidth]{{{./Phi0_8000}}}
    \includegraphics[width=0.16\linewidth]{{{./Phi0_10000}}}
    \includegraphics[width=0.16\linewidth]{{{./Phi0_12000}}}
    \caption{Three-dimensional view of the bubble shape within liquid column.}
  \end{subfigure}

  \vspace{0.25cm}
  \begin{subfigure}[t]{\linewidth}
    \begin{overpic}[width=0.16\linewidth]{{{./Phi0S_2000}}} \put(7,3){\small $t=2000 \delta_t$} \end{overpic}
    \begin{overpic}[width=0.16\linewidth]{{{./Phi0S_4000}}} \put(7,3){\small $t=4000 \delta_t$}\end{overpic}
    \begin{overpic}[width=0.16\linewidth]{{{./Phi0S_6000}}} \put(7,3){\small $t=6000 \delta_t$} \end{overpic}
    \begin{overpic}[width=0.16\linewidth]{{{./Phi0S_8000}}} \put(7,3){\small $t=8000 \delta_t$} \end{overpic}
    \begin{overpic}[width=0.16\linewidth]{{{./Phi0S_10000}}} \put(7,3){\small $t=10000 \delta_t$} \end{overpic}
    \begin{overpic}[width=0.16\linewidth]{{{./Phi0S_12000}}} \put(7,3){\small $t=12000 \delta_t$} \end{overpic}
    \caption{Slice through the center of the domain ($y=250 \delta_x$).}
  \end{subfigure}

  \caption{\label{fig:Phi0}Bubble rise without particles, at selected time steps
    (ordered from left to right). The spherical cap shape agrees well with the
    prediction according to \cite{Grace73}.}
\end{figure}
\begin{figure}
  \centering

  \begin{subfigure}[t]{\linewidth}
    \includegraphics[width=0.16\linewidth]{{{./Phi2.5_2000}}}
    \includegraphics[width=0.16\linewidth]{{{./Phi2.5_4000}}}
    \includegraphics[width=0.16\linewidth]{{{./Phi2.5_6000}}}
    \includegraphics[width=0.16\linewidth]{{{./Phi2.5_8000}}}
    \includegraphics[width=0.16\linewidth]{{{./Phi2.5_10000}}}
    \includegraphics[width=0.16\linewidth]{{{./Phi2.5_12000}}}
    \caption{Three-dimensional view of bubbles and particles. Particle color
      indicates initial $z$ coordinate of particle position at $t=0$.}
  \end{subfigure}

  \vspace{0.25cm}
  \begin{subfigure}[t]{\linewidth}
    \begin{overpic}[width=0.16\linewidth]{{{./Phi2.5S_2000}}}  \end{overpic}
    \begin{overpic}[width=0.16\linewidth]{{{./Phi2.5S_4000}}} \end{overpic}
    \begin{overpic}[width=0.16\linewidth]{{{./Phi2.5S_6000}}}  \end{overpic}
    \begin{overpic}[width=0.16\linewidth]{{{./Phi2.5S_8000}}}  \end{overpic}
    \begin{overpic}[width=0.16\linewidth]{{{./Phi2.5S_10000}}}  \end{overpic}
    \begin{overpic}[width=0.16\linewidth]{{{./Phi2.5S_12000}}}  \end{overpic}
    \caption{Slice through the center of the domain ($y=250 \delta_x$).}
  \end{subfigure}

  \caption{\label{fig:Phi2.5}Bubble rise from a bed consisting of $44\,621$
    particles ($\Phi=2.5\%$) at selected time steps (times chosen identical to
    \cref{fig:Phi0}).}
\end{figure}
\begin{figure}
  \centering

  \begin{subfigure}[t]{\linewidth}
    \includegraphics[width=0.16\linewidth]{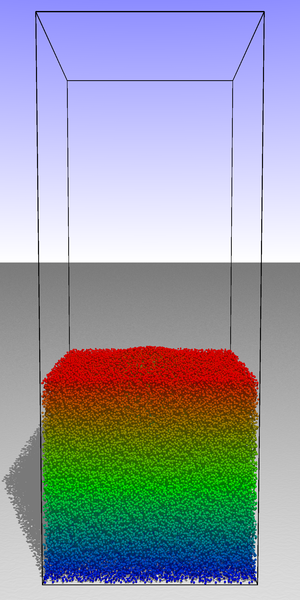}
    \includegraphics[width=0.16\linewidth]{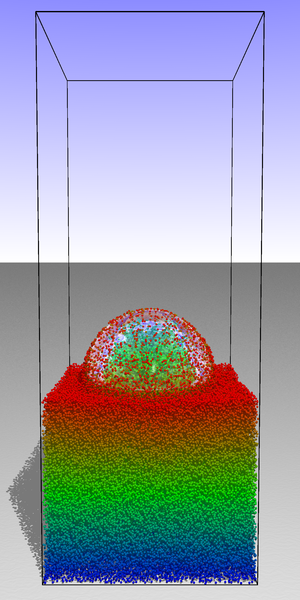}
    \includegraphics[width=0.16\linewidth]{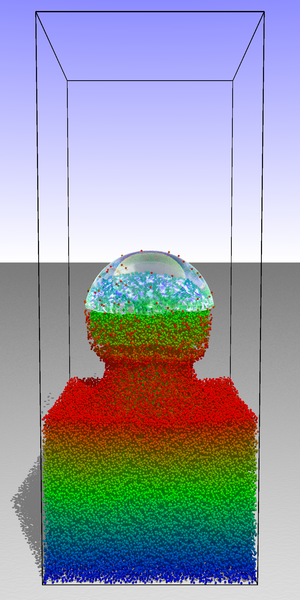}
    \includegraphics[width=0.16\linewidth]{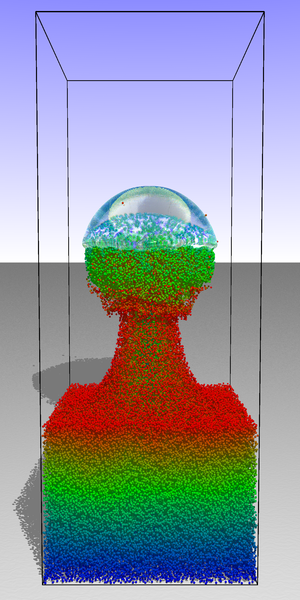}
    \includegraphics[width=0.16\linewidth]{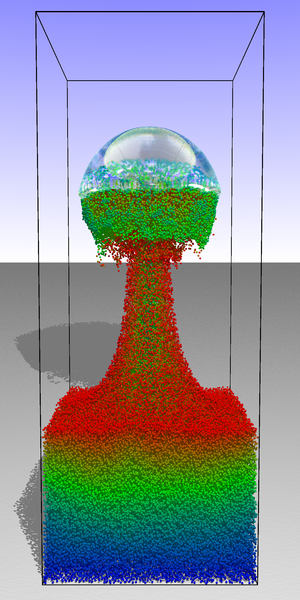}
    \includegraphics[width=0.16\linewidth]{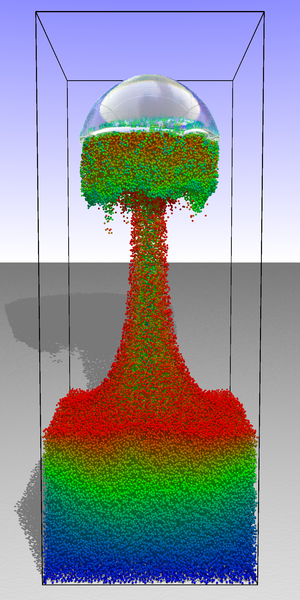}
    \caption{Three-dimensional view of bubbles and particles. Particle color
      indicates initial $z$ coordinate of particle position at $t=0$.}
  \end{subfigure}

  \vspace{0.25cm}
  \begin{subfigure}[t]{\linewidth}
    \begin{overpic}[width=0.16\linewidth]{{{./Phi10S_2000}}}  \end{overpic}
    \begin{overpic}[width=0.16\linewidth]{{{./Phi10S_4000}}} \end{overpic}
    \begin{overpic}[width=0.16\linewidth]{{{./Phi10S_6000}}}  \end{overpic}
    \begin{overpic}[width=0.16\linewidth]{{{./Phi10S_8000}}}  \end{overpic}
    \begin{overpic}[width=0.16\linewidth]{{{./Phi10S_10000}}}  \end{overpic}
    \begin{overpic}[width=0.16\linewidth]{{{./Phi10S_12000}}}  \end{overpic}
    \caption{Slice through the center of the domain ($y=250 \delta_x$).}
  \end{subfigure}

  \caption{\label{fig:Phi10}Bubble rise from a bed consisting of $178\,486$
    particles ($\Phi=10\%$) at selected time steps (times chosen identical to
    \cref{fig:Phi0}).}
\end{figure}

Without any particles ($\Phi = 0$), the terminal rise Reynolds number obtained
from simulations is $\Rey_b^*=48$. Again, the lower velocity can be attributed
to wall effects. \Cref{fig:Phi0,fig:Phi2.5,fig:Phi10} show the process without
particles and at bed densities $\Phi=2.5 \%$ and $\Phi=10 \%$.  The terminal
rise velocity is hardly affected by the presence of the particles in this case,
after the top of the bubble is uncovered from particles. In these images, the
particle color indicates the initial $z$ - position of the particle, to show the
mixing of different bed layers in the wake of the bubble. A circulating motion
is observable in the wake of the bubble. This effect seems to be absent in the
unresolved simulations of \citet{Deen2007BP}\new{, where the recirculation region
seems to have no influence on the particles.} Also a large number of particles
from the middle and lower layer of the bed are carried in the wake of the
bubble. That is, we can observe substantial changes in the relative positions of
particles during the mixing process.

\begin{figure}
  \centering
   \begin{subfigure}[t]{0.24\linewidth}
    \centering
    \includegraphics[width=\linewidth]{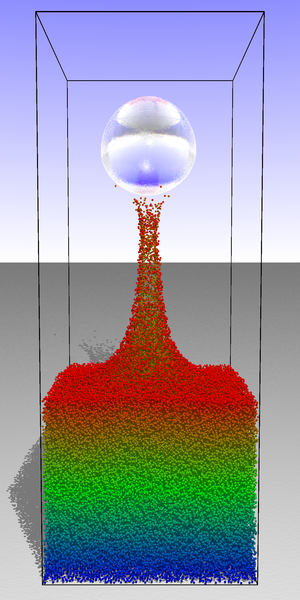}
    \caption{spherical, $\Eo=2.71$, $\Mo=1.26 \times 10^{-3}$}
  \end{subfigure}
  \begin{subfigure}[t]{0.24\linewidth}
    \centering
    \includegraphics[width=\linewidth]{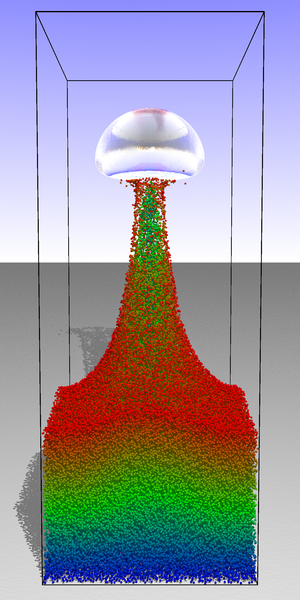}
    \caption{ellipsoidal, $\Eo=27.1$, $\Mo=9.09 \times 10^{-2}$}
  \end{subfigure}
  \begin{subfigure}[t]{0.24\linewidth}
    \centering
    \includegraphics[width=\linewidth]{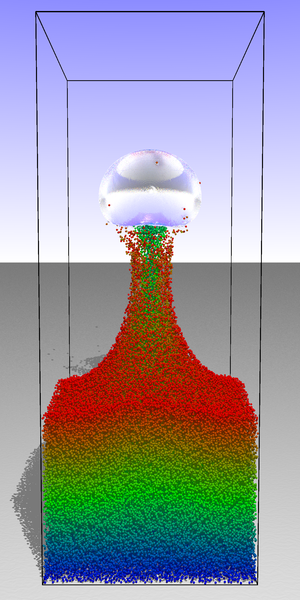}
    \caption{dimpled, $\Eo=271$, $\Mo=1.02 \times 10^{3}$}
  \end{subfigure}
  \begin{subfigure}[t]{0.24\linewidth}
    \centering
    \includegraphics[width=\linewidth]{./Phi10_10000}
    \caption{skirted, $\Eo=39.2$, $\Mo=9.81 \times 10^{-4}$}
  \end{subfigure}
  \caption{\label{fig:wakes}Particles mixing in the wake of bubbles at different
    bubble regimes (initial bed density $\Phi=10\%$). The simulations allow a
    detailed investigation of the wake structure and the accompanying mixing
    process of the particles. In the spherical case (a), only a thin filament of
    particles follows the upward motion of the bubble. Hardly any particles from
    lower layers (green or blue) of the bed are carried upwards. In the
    ellipsoidal case (b), the effect is stronger and one observes a significant
    portion of green particles following the bubble wake. Cases (c) and (d)
    feature a circulating flow in the wake of the bubble. This recirculating
    region in the wake of the bubble can carry a larger number of particles. The
    effect is strongest in case (d), where particles from lower layers of the
    bed (green) change relative position with particles from the top (red).}
\end{figure}

\Cref{fig:wakes} shows the same setup with altered fluid properties
corresponding to different bubble regimes. The wake and particle structure
differs significantly in the different regimes. The bubble wake is strongest for
the skirted regime that generates the largest recirculation region that also
displays the highest solid mass transport. At the other extreme, we find that
the spherical case has the least mixing effect and generates only a thin cone of
lifted particles.

Scenarios of this complexity require a considerable amount of computational
cost. The run time in each of the above cases is approximately $6.66$ hours when
utilizing $2\, 000$ cpu cores of a distributed system in parallel. \new{The
  considerable complexity stems from the high resolutions required by the
  DNS. Here, we have used $3.25 \cdot 10^8$ grid points, whereas the unresolved
  simulations of \citet{Deen2007BP} require only $1.6 \cdot 10^5$ grid points.}


\section{\label{sec:conclusion}Conclusion}
Direct numerical simulation of liquid-gas-solid flows offers the possibility of
detailed studies of bubble-particle interaction in liquids. The parallel model
proposed in this paper allows, to the best of our knowledge, for the first time
particle-resolved simulations of gas bubbles within slurry columns. This is
possible thanks to the parallel design of the model that allows the exploitation
of the parallelism of modern supercomputers. It has been demonstrated that the
model can simulate particle mixing in the wake of rising gas bubbles. The
structures formed by particles in the wake of the bubbles can be studied in
great detail. The effect of different bubble regimes, i.e., bubble size and
surface tension, on the particle transport can be analyzed. \new{A comparison
  with previous, unresolved simulations indicates that particle-resolved DNS is
  indeed necessary to predict this flow structure correctly.}

Future research may include particle wettability and structures at liquid-gas
interfaces. Also, systematic studies of bubble-particle interactions could be
used in improving existing drag correlations for bubbles in particle solutions.



\bibliographystyle{gCFD}

\bibliography{literature}

\end{document}